\documentclass[12pt]{article}
\usepackage{epsfig}
\usepackage{pst-plot,url,epsf}
\newcommand{\beq}{\begin{equation}}
\newcommand{\eeq}{\end{equation}}
\newcommand{\bea}{\begin{eqnarray}}
\newcommand{\eea}{\end{eqnarray}}
\newcommand{\beas}{\begin{eqnarray*}}
\newcommand{\eeas}{\end{eqnarray*}}

\newcommand{\epm}{e^+e^-}

\newcommand{\ra}{\rightarrow}
\newcommand{\ga}{\gamma}

\newcommand{\ttbar}{t\bar{t}}

\newcommand{\eebnmbdu}{e^+ e^- \ra b \nu_{\mu} \mu^+ \bar{b} d \bar{u}}
\newcommand{\eebnmbmn}{e^+ e^- \ra b \nu_{\mu} \mu^+ \bar{b} \mu^- 
                       \bar{\nu}_{\mu}}

\newcommand{\AmS}{{\protect\the\textfont2
  A\kern-.1667em\lower.5ex\hbox{M}\kern-.125emS}}
\newcommand{\eebudbdu}{e^+ e^- \ra b u \bar{d} \bar{b} d \bar{u}}

\newcommand{\nn}{\nonumber}
\def \uo{\ensuremath { u(p_2) \,\, }}

\def \vt{\ensuremath { \bar{v}(p_1) \,\, }}
\def\unity{{\rm 1\mskip-4.25mu l}}
\def \ks {k\hspace{-0.43em}/}

\newcommand{\fh}[2]{\hat{F}_{#1}^{#2}}

\begin{document}
\thispagestyle{empty}
\begin{flushright}
DESY 05-199\\
SFB/CPP-05-68\\
October 2005\\
\vspace*{1.5cm}
\end{flushright}
\begin{center}
{\LARGE\bf Factorizable electroweak $\cal{O}(\alpha)$ corrections
           for top quark pair production and decay
           at\\[1mm] a linear $\epm$ collider\footnote{Work supported in
           part by 
           the Polish State Committee for Scientific Research in years 
           2005--2007 as a research grant, by the European
           Community's Human Potential Program under contract
           HPRN-CT-2000-00149 Physics at Colliders and
           by DFG under Contract SFB/TR 9-03.}}\\
\vspace*{2cm}
Karol Ko\l odziej,\footnote{E-mails: kolodzie@us.edu.pl,
staron@server.phys.us.edu.pl} Adam Staro\'n$^2$\\[6mm]
{\small\it Institute of Physics, University of Silesia\\ 
ul. Uniwersytecka 4, PL-40007 Katowice, Poland}\\[8mm]
Alejandro Lorca\footnote{E-mails: alorca@ifh.de, Tord.Riemann@desy.de}
and Tord Riemann$^3$\\[6mm]
{\small\it Deutsches Elektronen-Synchrotron DESY\\
         Platanenallee 6, D-15738 Zeuthen, Germany}\\
\vspace*{2.5cm}
{\bf Abstract}\\
\end{center}
We calculate the standard model predictions for top quark pair
production and decay into six fermions at a linear $e^+e^-$ collider.
We include the factorizable electroweak $\cal{O}(\alpha)$ corrections in the
pole approximation and QED corrections due to the initial state radiation
in the structure function approach. The effects of the radiative corrections
on the predictions are illustrated by showing numerical results for two
selected six-fermion reactions $\eebnmbmn$ and $\eebnmbdu$.
\vfill
\newpage

\section{INTRODUCTION}
Precise measurements of top quark pair production
\bea
\label{eett}
         \epm \ra t \bar{t}
\eea
at the threshold and in the continuum region will
belong to the basic
physics program of the future International Linear Collider (ILC) \cite{ILC}.
In order to fully profit from these high precision measurements one has to
bring theoretical predictions to at least the same, or preferably
better, precision, which obviously requires taking into account radiative
corrections. The latter should be calculated not only for the on-shell
production process (\ref{eett}). Due to their large widths
the $t$- and $\bar{t}$-quark of reaction (\ref{eett})  almost immediately
decay into $bW^+$ and $\bar{b}W^-$, respectively, and the $W$-bosons
subsequently into 2 fermions each, thus constituting six-fermion reactions of
the form
\beq
\label{ee6f}
  e^+e^-\;\; \ra \;\; bf_1\bar{f'_1} \bar{b}f_2 \bar{f'_2},
\eeq
where $f_1, f'_2 =\nu_{e}, \nu_{\mu}, \nu_{\tau}, u, c$ and 
$f'_1, f_2 = e^-, \mu^-, \tau^-, d, s$. Typical lowest order Feynman diagrams
of reaction (\ref{ee6f}) are shown in Fig.~1.

\begin{figure}[htb]
\vspace{140pt}
\includegraphics{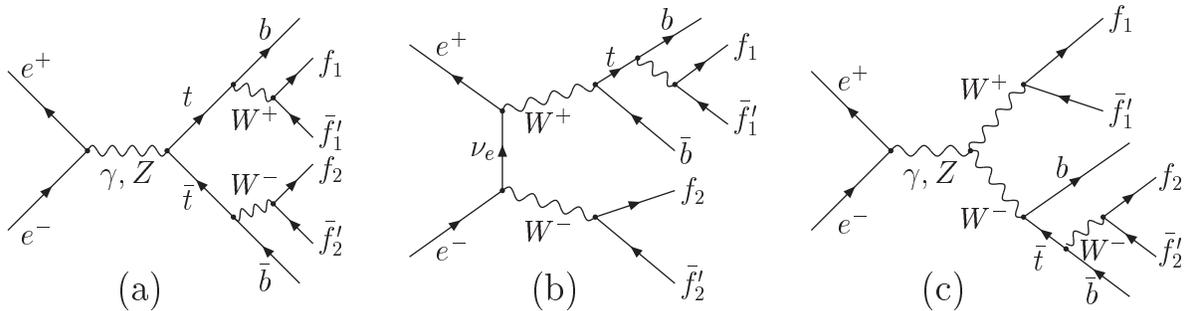}
\caption{Examples of Feynman diagrams of reaction (\ref{ee6f}):
(a) `signal', (b) and (c) `background'
diagrams.}
\label{fig:diags0}
\end{figure}

As decays of the top and antitop take place before toponium resonances 
can form, the Standard Model (SM) predictions for reaction (\ref{eett})
can be obtained with the perturbative method. The QCD predictions for reaction 
(\ref{eett}) in the threshold region were obtained in \cite{topQCD}
and then improved by calculation of the next-to-next-to-leading order QCD 
corrections \cite{topNNLO}, and by including the effects of initial state 
radiation and beamstrahlung \cite{topIR}.
The $\mathcal{O}(\alpha\alpha_s)$ \cite{topdec1,topdDS,topdD} and
$\mathcal{O}(\alpha\alpha_s^2)$ \cite{topdec2} corrections to the subsequent
top decay into a $W$ boson and a $b$ quark are also known.
In the continuum above the threshold, the QCD predictions for reaction 
(\ref{eett}) are known to order $\alpha_s^2$ \cite{eettQCD} and the
electroweak (EW) corrections to one--loop order \cite{eettEW,topfit1,topfit3},
including the hard bremsstrahlung
corrections \cite{eetthb,topfit1}. The QCD and EW corrections are
large, typically of $\cal{O}$(10\%). Order $\alpha_s$ \cite{eettQCD1}
and $\alpha_s^2$ QCD, and EW corrections have been combined in \cite{eettcomb}.
Quite recently the EW radiative corrections to (\ref{eett}) have been 
recalculated with a program {\tt topfit}
\cite{topfit1,topfit3} and thoroughly compared with results
of other calculations, with hard bremsstrahlung \cite{Fleischer:2002nn} and 
without it \cite{Hahn:2003ab}.
Finally, the radiative corrections to $W$ decays into fermion pairs, which
have to be taken into account too, are also 
known \cite{Bardin:1986fi,FJ2,Denner:1990tx}.

At tree level, reactions (\ref{ee6f}) can be studied with a
Monte Carlo (MC) program {\tt eett6f} \cite{eett6f,eett6f1} or with any other
MC program dedicated to the six fermion reactions, such as 
{\sc Sixphact}~\cite{Sixphact}, {\sc Sixfap}~\cite{Sixfap},
{\sc Lusifer} \cite{Lusifer},  or with any of 
multi-purpose generators,  such as {\sc Amegic} \cite{Amegic},
{\sc Grace} \cite{Grace}/{\sc Bases} \cite{Bases}, 
{\sc Madgraph} \cite{Madgraph}/{\sc Madevent} \cite{Madevent},
{\sc Phegas} \cite{Phegas}/{\sc Helac} \cite{Helac}, 
or {\sc Whizard} \cite{Whizard}/{\sc Comphep}~\cite{Comphep}, 
{\sc Madgraph}~\cite{Madgraph}, or {\sc O'mega}~\cite{Omega}.
Thorough comparison of the lowest order predictions for several
different channels of (\ref{ee6f})
obtained with {\sc Amegic++}, {\tt eett6f}, {\sc Lusifer}, {\sc Phegas},
{\sc Sixfap} and {\sc Whizard} have been performed in the framework of
the Monte Carlo Generators group of the ECFA/DESY workshop \cite{didi}.
A survey of SM cross sections of all six fermion reactions 
with up to four quarks
in the limit of 
massless fermions (but the top quark), has been done
in \cite{Lusifer}. The latter contains also a fine tuned comparison of both
the lowest order and lowest order plus ISR results, obtained in the
structure function approach, between {\sc Lusifer} and {\sc Whizard}.

Concerning radiative corrections to the six-fermion reactions (\ref{ee6f}),
the situation is less advanced.
Already at the tree level, any of the reactions receives
contributions from typically several hundred Feynman diagrams, {\em e.g.}
in the unitary gauge, with neglect of the Higgs boson couplings to
fermions lighter than the $b$ quark, reactions $\eebnmbdu$, $\eebnmbmn$,
and $\eebudbdu$ get contributions from 264, 452, and 1484 Feynman diagrams,
respectively. Hence, the calculation 
of the full $\cal{O}(\alpha)$ radiative corrections to any of reactions 
(\ref{ee6f}) seems not to be feasible at present.
Therefore, in the present note we will make a step
towards improving precision of the lowest order predictions for
(\ref{ee6f}) by  including leading radiative effects, such as
initial state radiation (ISR) and factorizable EW radiative 
corrections to the process of the on-shell top quark pair production
(\ref{eett}), to the decay of the $t$ ($\bar t$) into $bW^+$ ($\bar{b}W^-$) and
to the subsequent decays of the $W$-bosons. 
We will illustrate an effect of of these corrections by showing numerical 
results for the two selected six-fermion reactions
\beq
\label{nmmn}
\eebnmbmn
\eeq
and
\beq
\label{nmud}
\eebnmbdu.
\eeq

\section{CALCULATIONAL SCHEME}

We calculate the ISR and the factorizable SM corrections
for the reaction
\beq
\label{ee6fmom}
  e^+(p_1,\sigma_1)\;e^-(p_2,\sigma_2)\;\; \ra \;\; b(p_3,\sigma_3)\;
f_1(p_4,\sigma_4)\; \bar{f'_1}(p_5,\sigma_5)\; \bar{b}(p_6,\sigma_6) \;
f_2(p_7,\sigma_7) \; \bar{f'_2}(p_8,\sigma_8),
\eeq
where the particle momenta and helicities have been indicated in the 
parentheses, according 
to the following formula:
\bea
\label{LL}
{\rm d} \sigma=
\int_0^1 {\rm d} x_1 \int_0^1 {\rm d} x_2 \,
          \Gamma_{ee}^{LL}\left(x_1,Q^2\right)
\Gamma_{ee}^{LL}\left(x_2,Q^2\right)
{\rm d}\sigma_{\rm Born+FEWC}\left(x_1 p_1,x_2 p_2\right),
\eea
where $x_1p_1$ ($x_2p_2$) is the four momentum of the positron
(electron) after emission of a collinear photon.
The structure function $\Gamma_{ee}^{LL}\left(x,Q^2\right)$ is given
by Eq.~(67) of \cite{Beenakker}, with {\tt `BETA'} choice for non-leading
terms. The splitting scale $Q^2$, which is not fixed in the LL approximation
is chosen to be $s=(p_1+p_2)^2$.
By ${\rm d}\sigma_{\rm Born+FEWC}$ we denote the cross section including
the factorizable EW $\cal{O}(\alpha)$ corrections
\bea
\label{bpfewc}
{\rm d}\sigma_{\rm Born+FEWC}
=\frac{1}{2s}\left\{\overline{\left|M_{\rm Born}\right|^2}\;
+ 2\;{\rm Re}\overline{\left(M_{t\bar{t}}^{*}
\;\delta M_{t\bar{t},{\rm FEW}}\right)}\right\}{\rm d}\Phi_{6f},
\eea
where $M_{\rm Born}$ is the matrix element of reaction (\ref{ee6fmom})
obtained with the complete set of the lowest order Feynman diagrams,
$M_{t\bar{t}}$ and $\delta M_{t\bar{t},{\rm FEW}}$ is, respectively, 
the lowest order amplitude of the `signal' Feynman diagram of 
Fig.~\ref{fig:diags0}a and the corresponding factorizable EW 
$\cal{O}(\alpha)$ correction, both in the pole approximation.
The overlines in (\ref{bpfewc}) denote, as usual, an initial state particle 
spin average and a sum over final state particle polarizations,
and ${\rm d}\Phi_{6f}$ is the Lorentz invariant six-particle 
phase space element. The basic phase space parametrizations which are used 
in the program are given by Eqs.~(7)--(9) of \cite{eett6f}.
The corrections 
that we take into account in $\delta M_{t\bar{t},{\rm FEW}}$  
are illustrated diagramatically in Fig.~2. 

\begin{figure}[htb]
\vspace{240pt}
\includegraphics{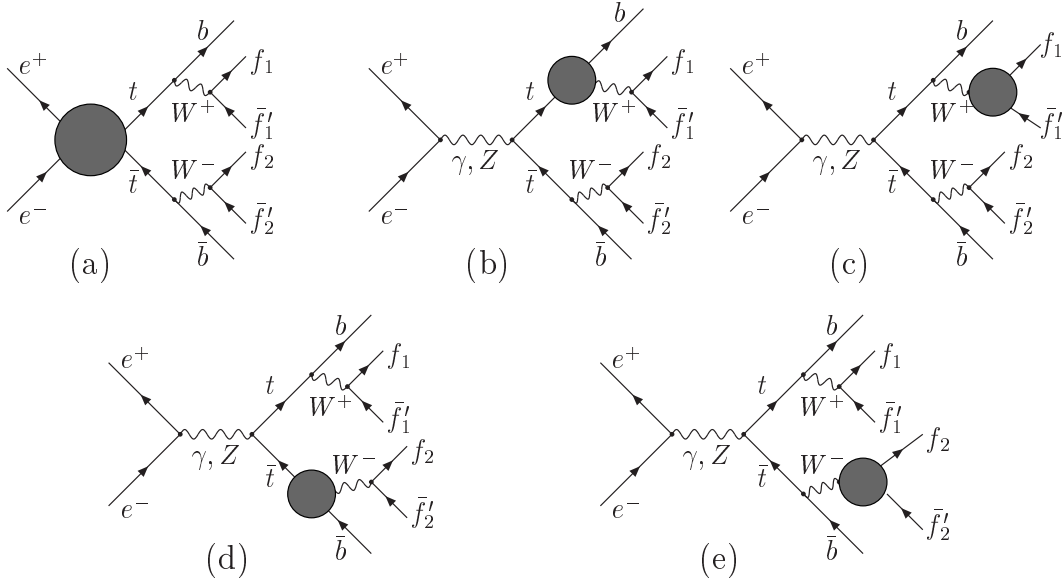}
\caption{Factorizable EW corrections to reaction (\ref{ee6f}).}
\label{fig:diags}
\end{figure}

In the pole approximation, the polarized lowest order amplitude 
$M_{t\bar{t}}$ and the one--loop correction $\delta M_{t\bar{t},{\rm FEW}}$  
of Eq.~(\ref{bpfewc}) can be expressed analytically as follows:
\bea
M_{t\bar{t}}^{\sigma_1\sigma_2;\sigma_3\ldots\;\sigma_8}=
\frac{1}{D_t\left(p_{345}\right)D_t\left(p_{678}\right)}
   \sum_{\sigma_t, \sigma_{\bar t}}& &\hspace*{-0.5cm}
M_{e^+e^-\ra t\bar{t}}^{\sigma_1\sigma_2;\sigma_t \sigma_{\bar t}}\;
M_{t\ra b f_1 f'_1}^{\sigma_t;\sigma_3\sigma_4\sigma_5}\;
M_{\bar{t}\ra \bar{b} f_2f'_2}^{\sigma_{\bar{t}};\sigma_6\sigma_7\sigma_8} 
\label{mtt}\\
\delta M_{t\bar{t}}^{\sigma_1\sigma_2;\sigma_3\ldots\;\sigma_8}=
\frac{1}{D_t\left(p_{345}\right)D_t\left(p_{678}\right)}
\sum_{\sigma_t, \sigma_{\bar t}}& &\hspace*{-0.5cm}\left[
\delta M_{e^+e^-\ra t\bar{t}}^{\sigma_1\sigma_2\sigma_t, \sigma_{\bar t}}
M_{t\ra b f_1 f'_1}^{\sigma_t;\sigma_3\sigma_4\sigma_5}\;
M_{\bar{t}\ra \bar{b} f_2f'_2}^{\sigma_{\bar{t}};\sigma_6\sigma_7\sigma_8}
\right.\nn\\
&+&\hspace*{-0.3cm}\left. 
M_{e^+e^-\ra t\bar{t}}^{\sigma_1\sigma_2;\sigma_t\sigma_{\bar t}}\;
\delta M_{t\ra b f_1 f'_1}^{\sigma_t\sigma_3\sigma_4\sigma_5}\;
M_{\bar{t}\ra \bar{b} f_2f'_2}^{\sigma_{\bar{t}};\sigma_6\sigma_7\sigma_8}
    \right.\label{dmtt}\\
&+&\hspace*{-0.3cm} \left. 
      M_{e^+e^-\ra t\bar{t}}^{\sigma_1\sigma_2;\sigma_t\sigma_{\bar t}}\;
      M_{t\ra b f_1 f'_1}^{\sigma_t;\sigma_3\sigma_4\sigma_5}\;
      \delta M_{\bar{t}\ra \bar{b} f_2f'_2}^{\sigma_{\bar{t}};
                        \sigma_6\sigma_7\sigma_8}\right],\nn
\eea
where the lowest order $t$ and $\bar t$ decay amplitudes and the corresponding
one--loop corrections read
\bea
M_{t\ra b f_1 f'_1}^{\sigma_t\sigma_3\sigma_4\sigma_5}&=&
\frac{1}{D_W\left(p_{45}\right)}\sum_{\lambda_{W^+}}
M_{t\ra b W^+}^{\sigma_t\sigma_3\lambda_{W^+}}
M_{W^+\ra f_1 f'_1}^{\lambda_{W^+}\sigma_4\sigma_5},
\label{mtbff}\\
M_{\bar{t}\ra \bar{b} f_2f'_2}^{\sigma_{\bar{t}}\sigma_6\sigma_7\sigma_8}
&=&\frac{1}{D_W\left(p_{78}\right)}\sum_{\lambda_{W^-}}
M_{\bar{t}\ra \bar{b} W^-}^{\sigma_{\bar{t}}\sigma_6\lambda_{W^-}}
      M_{W^-\ra f_2 f'_2}^{\lambda_{W^-}\sigma_7\sigma_8},
\label{mtbbff}\\
\delta M_{t\ra b f_1 f'_1}^{\sigma_t\sigma_3\sigma_4\sigma_5}&=&
\frac{1}{D_W\left(p_{45}\right)}\sum_{\lambda_{W^+}}\left[
\delta M_{t\ra b W^+}^{\sigma_t\sigma_3\lambda_{W^+}}
      M_{W^+\ra f_1 f'_1}^{\lambda_{W^+}\sigma_4\sigma_5}
     +M_{t\ra b W^+}^{\sigma_t\sigma_3\lambda_{W^+}}
      \delta M_{W^+\ra f_1 f'_1}^{\lambda_{W^+}\sigma_4\sigma_5}\right],
\label{dmtbff}\\
\delta M_{\bar{t}\ra \bar{b} f_2f'_2}^{\sigma_{\bar{t}}
\sigma_6\sigma_7\sigma_8}&=&\frac{1}{D_W\left(p_{78}\right)}
   \sum_{\lambda_{W^-}}\left[
\delta M_{\bar{t}\ra \bar{b} W^-}^{\sigma_{\bar{t}}\sigma_6\lambda_{W^-}}
      M_{W^-\ra f_2 f'_2}^{\lambda_{W^-}\sigma_7\sigma_8}
     +M_{\bar{t}\ra \bar{b} W^-}^{\sigma_{\bar{t}}\sigma_6\lambda_{W^-}}
      \delta M_{W^-\ra f_2 f'_2}^{\lambda_{W^-}\sigma_7\sigma_8}\right].
\label{dmtbbff}
\eea
In (\ref{mtt}--\ref{dmtbbff}), 
$\sigma_t$, $\sigma_{\bar t}$ and $\lambda_{W^+}$, $\lambda_{W^-}$
denote polarizations of the intermediate top quarks and $W$ bosons
which are treated as on-shell particles, except for keeping their actual 
off-shell momenta
\beq
p_{345}=p_3+p_4+p_5, \quad p_{678}=p_6+p_7+p_8, \qquad p_{78}=p_7+p_8,
\quad p_{45}=p_4+p_5
\eeq
in the denominators $D_t\left(p\right)$ and $D_W\left(p\right)$ of their
propagators
\beq
\label{props}
D_t\left(p\right)=p^2 - m_t^2 + i m_t\Gamma_t, \qquad
D_W\left(p\right)=p^2 - m_W^2 + i m_W\Gamma_W.
\eeq
The fixed widths $\Gamma_t$ and $\Gamma_W$ of (\ref{props}) are calculated in
the program for a given set of initial parameters. They are set to
their SM lowest order values, $\Gamma_t^{(0)}$ and $\Gamma_W^{(0)}$, 
for the Born cross sections, or they
include radiative corrections of the same kind as those included in the 
numerators of (\ref{dmtt}), (\ref{dmtbff}) and (\ref{dmtbbff}) for the
radiatively corrected cross sections.

While explaining further the notation of Eqs. (\ref{mtt}--\ref{dmtbbff}) we
will suppress the polarization indices.
$M_{e^+e^-\ra t\bar{t}}$ and $\delta M_{e^+e^-\ra t\bar{t}}$
are the lowest order and the EW one--loop amplitudes
of the on-shell top quark pair production process (\ref{eett}). They can be
decomposed in a basis composed of the following invariant amplitudes
\bea
\label{amps}
{\cal M}_{1,\,{ab}} & =& \vt \ga^{\mu}\, g_a \, \uo \; \bar{u}(k_t)
\ga_{\mu}\, g_b \, \bar{v}(k_{\bar t}), 
\quad  g_a, g_b=\unity, \gamma_5,\nonumber\\
{\cal M}_{3,{11}} & = &-\vt \ks_t \, \uo \;\bar{u}(k_t) 
                                              \bar{v}(k_{\bar t}),\\
{\cal M}_{3,{51}} & = &-\vt \ks_t\, \ga_5  \uo \;
\bar{u}(k_t) \bar{v}(k_{\bar t}).\nonumber
\eea
The projected four momenta $k_t, k_{\bar t}$ of the on-shell
top- and antitop-quark of (\ref{amps}), as well as the four momenta 
$k_{W^+}, k_{W^-}$ of the on-shell $W$-bosons and the four momenta 
$k_3,\ldots,k_8$ of the decay fermions, which are used later,
have been obtained
from the four momenta of the final state fermions $p_3,\ldots,p_8$ of reaction
(\ref{ee6f}) with
the projection procedure described in Appendix A. 

In terms of invariant amplitudes (\ref{amps}), the lowest order amplitude
of (\ref{eett}) reads
\bea
\label{Bornamp}
 \, M_{e^+e^-\ra t\bar{t}} = 
\sum_{a,b=1,5} {\rm F}_{1B}^{\,ab} \,\,    {\cal M}_{1,{ab}},
\eea
where the 4 Born form factors ${\rm F}_{1B}^{\,ab}$ are given by
\\
\parbox{8cm}{
\beas
{\rm F}_{1B}^{\,11}&=&\frac{e_W^2\left(\chi_Z v_e v_t+Q_e Q_t\right)}{s},\\
{\rm F}_{1B}^{\,15}&=&-\frac{e_W^2\chi_Z v_e a_t}{s},
\eeas}
\hfill
\parbox{6cm}{
\beas
{\rm F}_{1B}^{\,51}&=&-\frac{e_W^2\chi_Z v_t a_e}{s},\\
{\rm F}_{1B}^{\,55}&=&\frac{e_W^2\chi_Z a_e a_t}{s}.
\eeas}
\hfill
\parbox{2cm}{\bea \label{fborn}  \eea}\\
In (\ref{fborn}), $e_W$ is the effective electric charge, 
$e_W=\sqrt{4\pi\alpha_W}$, with
\beq
\label{sw2}
\alpha_W=\frac{\sqrt{2}G_{\mu}m_W^2 \sin^2\theta_W}{\pi}
\qquad {\rm and} \qquad \sin^2\theta_W=1-\frac{m_W^2}{m_Z^2},
\eeq
the $Z$-boson propagator is contained in the factor
\beq
\label{chiz}
\chi_Z=\frac{1}{4 \sin^2\theta_W \cos^2\theta_W}\;
\frac{s}{s-m_Z^2+im_Z\Gamma_Z}
\eeq
and we have used the following conventions for couplings of the electron 
and top quark to a photon and $Z$-boson
\beq
Q_e=-1, \quad Q_t=\frac{2}{3}, \quad a_e=-a_t=-\frac{1}{2}, \quad 
v_f=a_f\left(1-4\left|Q_f\right|\sin^2\theta_W\right), \quad f=e,t.
\eeq
We have introduced a constant $Z$-boson width $\Gamma_Z$ in (\ref{chiz}),
in a similar way as $\Gamma_t$ and $\Gamma_W$ have been introduced in
(\ref{props}), although the $Z$-boson propagator in the $\epm$ annihilation 
channel never becomes resonant in the CMS energy range above the 
$t\bar t$-pair production threshold. Generally speaking, 
the constant width
$\Gamma$ of an unstable particle is introduced into the lowest order
matrix elements by replacing its mass with the complex mass parameter
\beq
\label{m2}
m^2 \ra m^2-im\Gamma
\eeq
in the corresponding propagator, both in the $s$- and $t$-channel one,
while keeping the electroweak mixing parameter $\sin^2\theta_W$ of
(\ref{sw2}) real.
This approach is usually referred to in the literature as the fixed width
scheme (FWS). The approach, in which $m_W^2$ and $m_Z^2$ are replaced with 
their complex counterparts according to (\ref{m2}) also in $\sin^2\theta_W$ of 
(\ref{sw2}) is on the other hand referred to as the complex mass scheme
\cite{Racoon}. The latter has the advantage that it
preserves Ward identities. Let us note, that in 
Eqs.~(\ref{mtt}--\ref{dmtbbff}), substitution (\ref{m2}) is done
only in the denominators of the top-quark and $W$-boson propagators
and not in the one--loop amplitudes. Also the sums over the top-quark 
and $W$-boson polarizations result in the numerators of the corresponding 
propagators with real masses. However, 
this does not violate the substitution rule of (\ref{m2}),
as the amplitudes of Eqs.~(\ref{mtt}--\ref{dmtbbff}) constitute 
the factorizable one--loop correction term in (\ref{bpfewc}).

The EW one--loop amplitude of (\ref{eett}) reads
\bea
\label{damp}
 \delta M_{e^+e^-\ra t\bar{t}} = 
\sum_{a,b=1,5} \fh{1}{ab} \,\,    {\cal M}_{1,{ab}} +  
 \fh{3}{11} \,\,    {\cal M}_{3,{11}} +  
 \fh{3}{51} \,\,    {\cal M}_{3,{51}},
\eea
with the six independent form factors: $\fh{1}{ab}, \; a,b=1,5$, 
$\fh{3}{11}$ and $\fh{3}{51}$ which are calculated numerically with
a program {\tt topfit} \cite{topfit1,topfit3} that is tailored
to a subroutine of a new version of 
{\tt eett6f}. Note that a factor $i$ has been omitted
on the left hand side of (\ref{amps}) compared to \cite{topfit1}. 
Keeping it would result in an extra minus sign on the right hand 
side of (\ref{mtt}) and (\ref{dmtt}), as we
neglect the $i$ factor in every vertex and propagator 
and consequently the resulting common $+i$ factor for every Feynman
diagram in the present work.
The flags in {\tt topfit} switch off all photonic corrections there, 
including the running of the electromagnetic coupling. This means that only 
the genuine weak corrections will contribute.

In order to fix normalization we give the formula for the EW one--loop
corrected cross section ${\rm d}\sigma_{e^+e^-\ra\; t \bar t}$
of the on-shell top production (\ref{eett})
\bea
\label{cseett}
{\rm d}\sigma_{e^+e^-\ra t\bar{t}}
=\frac{1}{2s}\left\{\left|M_{e^+e^-\ra t\bar{t}}\right|^2\;
+ 2\;{\rm Re}\left(M_{e^+e^-\ra t\bar{t}}^{*}
\;\delta M_{e^+e^-\ra t\bar{t}}\right)\right\}{\rm d}\Phi_{2f},
\eea
where the matrix elements $M_{e^+e^-\ra t\bar{t}}$ and 
$\delta M_{e^+e^-\ra t\bar{t}}$ are given by (\ref{Bornamp}) and
(\ref{damp}) and ${\rm d}\Phi_{2f}$ is the Lorentz invariant two-particle 
phase space element
\beq
\label{dps2}
 {\rm d}\Phi_{2f} = \frac{|\vec{p_t}|}{4\sqrt{s}} {\rm d} \Omega_t,
\eeq
with $\vec{p_t}$ being the momentum and $\Omega_t$ the solid angle
of the $t$-quark.

The $t$- and $\bar t$-quark decay amplitudes $M_{t \ra b W^+}$ 
and $M_{\bar t \ra \bar b W^-}$,
and the corresponding one--loop corrections $\delta M_{t \ra b W^+}$ 
and $\delta M_{\bar t \ra \bar b W^-}$ can be decomposed in terms
of the invariant amplitudes\\
\parbox{8cm}{
\beas
{\cal M}_{t,1}^{(\sigma)} &=& \bar{u}(k_3) \varepsilon\!\!\!/(k_{W^+})
                                                    P_{\sigma}u(k_t),\\
{\cal M}_{t,2}^{(\sigma)} &=& k_t\cdot \varepsilon(k_{W^+})\;
                             \bar{u}(k_3) P_{\sigma}u(k_t).
\eeas}
\hfill
\parbox{6cm}{
\beas
{\cal M}_{\bar t,1}^{(\sigma)} &=& 
\bar{v}(k_{\bar t}) \varepsilon\!\!\!/(k_{W^-})P_{\sigma}v(k_6), \\
{\cal M}_{\bar t,2}^{(\sigma)} &=&-k_{\bar t}\cdot \varepsilon(k_{W^-})\;
                             \bar{v}(k_{\bar t}) P_{\sigma}v(k_6).
\eeas}
\hfill
\parbox{2cm}{\bea \label{invtop}  \eea},\\
where $P_{\sigma}=(1+\sigma\gamma_5)/2$, $\sigma=\pm 1$, are the chirality 
projectors and we have used real polarization vectors for $W$ bosons.
The decomposition reads\\
\parbox{8cm}{
\beas
M_{t \ra b W^+}&=&g_{Wff}\;{\cal M}_{t,1}^{(-)},\\
\delta M_{t \ra b W^+}&=&g_{Wff}\;
\sum_{i=1,2\atop\sigma=\pm 1} F_{t,i}^{(\sigma)}{\cal M}_{t,i}^{(\sigma)},
\eeas}
\hfill
\parbox{6cm}{
\beas
M_{\bar t \ra \bar b W^-}&=&g_{Wff}\;{\cal M}_{\bar t,1}^{(-)},\\
\delta M_{\bar t \ra \bar b W^-}&=&g_{Wff}\;
\sum_{i=1,2\atop\sigma=\pm 1} F_{\bar t,i}^{(\sigma)}
                                  {\cal M}_{\bar t,i}^{(\sigma)}.
\eeas}
\hfill
\parbox{2cm}{\bea \label{amptop}  \eea}\\
In (\ref{amptop}), $g_{Wff}$ is the SM $W$ boson coupling to 
fermions which, similarly to the Born form factors of (\ref{fborn}), is 
defined in terms of the effective electric charge $e_W$
\beq
g_{Wff}=-\frac{e_W}{\sqrt{2}\sin\theta_W},
\eeq
$F_{t,i}^{(\sigma)}$ and $F_{\bar t,i}^{(\sigma)}$ are the EW one--loop
form factors of the top- and antitop-quark decay, respectively. The form 
factors $F_{t,i}^{(\sigma)}$ are calculated numerically with a newly written
dedicated subroutine that reproduces results of \cite{topdDS,topdD}.
The one--loop form factors of the antitop decay are then obtained  assuming
$\cal{C}\cal{P}$ conservation which lead to the following relations
\bea
F_{\bar t,1}^{(\sigma)}=F_{t,1}^{(\sigma)^*},\qquad
F_{\bar t,2}^{(\sigma)}=F_{t,2}^{(-\sigma)^*}.
\eea
Note that the imaginary parts of the form factors do not contribute 
at the one-loop order. 

Similarly the $W^+$- and $W^-$-boson decay amplitudes 
$M_{W^+ \ra f_1 \bar{f'}_1}$ and $M_{W^- \ra f_2 \bar{f'}_2}$,
and the corresponding one--loop corrections 
$\delta M_{W^+ \ra f_1 \bar{f'}_1}$ and $\delta M_{W^- \ra f_2 \bar{f'}_2}$ 
are given by\\
\parbox{8cm}{
\beas
M_{W^+ \ra f_1 \bar{f'}_1}&=&g_{Wff}\;{\cal M}_{W^+,1}^{(-)},\\
\delta M_{W^+ \ra f_1 \bar{f'}_1}&=&g_{Wff}\;
\sum_{i=1,2\atop\sigma=\pm 1} F_{W^+,i}^{(\sigma)}{\cal M}_{W^+,i}^{(\sigma)},
\eeas}
\hfill
\parbox{6cm}{
\beas
M_{W^- \ra f_2 \bar{f'}_2}&=&g_{Wff}\;{\cal M}_{\bar W^-,1}^{(-)},\\
\delta M_{W^- \ra f_2 \bar{f'}_2}&=&g_{Wff}\;
\sum_{i=1,2\atop\sigma=\pm 1} F_{W^-,i}^{(\sigma)}
                                  {\cal M}_{W^-,i}^{(\sigma)},
\eeas}
\hfill
\parbox{2cm}{\bea \label{ampw}  \eea}\\
with the invariant amplitudes\\
\parbox{8cm}{
\beas
{\cal M}_{W^+,1}^{(\sigma)} &=& \bar{u}(k_4) \varepsilon\!\!\!/(k_{W^+})
                                                    P_{\sigma}v(k_5),\\
{\cal M}_{W^+,2}^{(\sigma)} &=& k_4\cdot \varepsilon(k_{W^+})\;
                             \bar{u}(k_4) P_{\sigma}v(k_5).
\eeas}
\hfill
\parbox{6cm}{
\beas
{\cal M}_{W^-,1}^{(\sigma)} &=& \bar{u}(k_7) \varepsilon\!\!\!/(k_{W^-})
                                                    P_{\sigma}v(k_8), \\
{\cal M}_{W^-,2}^{(\sigma)} &=&-k_8\cdot \varepsilon(k_{W^-})\;
                             \bar{u}(k_7) P_{\pm}v(k_8)
\eeas}
\hfill
\parbox{2cm}{\bea \label{invw}  \eea}\\
and the EW one--loop form factors of the $W$-boson decays
$F_{W^{\pm},i}^{(\sigma)}$ being calculated numerically, this time with
a new subroutine that reproduces results of \cite{Denner:1990tx,topdD}
for the EW corrected $W$-boson width. Again, the imaginary parts of 
the form factors do not contribute 
at the one-loop order.

The calculation of the EW factorizable corrections to reaction (\ref{ee6f})
in the pole approximation makes sense only if the invariant masses
\bea
\label{mt}
m_{345}=\sqrt{\left(p_3+p_4+p_5\right)^2}, \qquad
m_{678}=\sqrt{\left(p_6+p_7+p_8\right)^2}
\eea
of the $bf_1\bar{f'_1}$ and $\bar{b}f_2 \bar{f'_2}$
are close to $m_t$ each and if
\bea
\label{mw}
m_{45}=\sqrt{\left(p_4+p_5\right)^2}, \qquad
m_{78}=\sqrt{\left(p_7+p_8\right)^2}
\eea
of the $f_1\bar{f'_1}$ and $f_2 \bar{f'_2}$ do not depart too much from $m_W$.
Otherwise the signal diagrams of Fig.~\ref{fig:diags0}(a) stop to dominate
the cross section and the
association of the reduced phase space point, at which the EW factorizable
$\cal{O}(\alpha)$ corrections depicted in Fig.~\ref{fig:diags} are calculated,
with the phase space point of the full six particle phase space of (\ref{ee6f})
may lead to unnecessary distortion of the off resonance background 
contributions.
Therefore in the following we will impose kinematical cuts on the quantities
\bea
\label{deltas}
\delta_t=m_{345}/m_t-1, \quad \delta_{\bar t}=m_{678}/m_t-1, \quad
         \delta_{W^+}=m_{45}/m_W-1,\quad \delta_{W^-}=m_{78}/m_W-1,
\eea
which describe the relative departures of the invariant masses
of (\ref{mt}) and (\ref{mw}) from $m_t$ and $m_W$, respectively.

\section{NUMERICAL RESULTS}

In this section, we will illustrate the effect of the factorizable EW
$\cal{O}(\alpha)$ corrections described in Section 2 on the SM predictions
for six fermion reactions relevant for detection of
the top quark pair production and decay at the ILC (\ref{ee6f})
by showing results for total cross sections of its two specific channels
(\ref{nmmn}) and (\ref{nmud}).

We choose the $Z$ boson mass, Fermi
coupling and fine structure constant in the Thomson limit
as the EW SM input parameters 
\bea
\label{params1}
m_Z=91.1876\; {\rm GeV},\qquad
G_{\mu}=1.16637 \times 10^{-5}\;{\rm GeV}^{-2}, \qquad
\alpha_0=1/137.0359895.
\eea
The external fermion masses of reaction (\ref{nmmn}) and the top quark mass
are the following:
\bea
\label{params3}
m_e=0.51099907\;{\rm MeV},\quad m_{\mu}=105.658389\;{\rm MeV},
\quad m_b=4.7\;{\rm GeV}, \quad m_t=178\;{\rm GeV}.
\eea
For definiteness, we give also values of the other fermion masses
\bea
\label{params4}
m_{\tau}=1.77705\;{\rm GeV}, \quad m_u=75\;{\rm MeV},
\quad m_d=75\;{\rm MeV},\quad m_s\!=250\;{\rm MeV},
\quad m_c\!=1.5\;{\rm GeV}
\eea
and the value of a strong coupling $\alpha_s(m_Z^2)=0.117$.

Assuming a value of the Higgs boson mass,
the $W$ boson mass and the $Z$ boson width are determined with
{\tt ZFITTER} \cite{ZFITTER}, while the SM Higgs boson width is calculated
with {\tt HDECAY} \cite{HDECAY}. We obtain the following values of these
parameters for $m_H=120$~GeV:
\bea
\label{params5}
m_W = 80.38509\; {\rm GeV}, \qquad \Gamma_Z = 2.495270\;{\rm GeV},
\qquad \Gamma_H=3.2780~{\rm MeV}.
\eea
The actual values of the $Z$ and Higgs boson widths are not very relevant 
in the context of the top quark pair production as they enter the calculation
through the off resonance background contributions.
The EW corrected top quark and $W$ boson widths, which on the 
other hand play
an essential role for the calculation, are calculated with a newly written
dedicated subroutine that reproduces results of \cite{Denner:1990tx,topdD}.
We obtain the following values for them for
the parameters specified in (\ref{params1}--\ref{params4})
\bea
\label{params6}
\Gamma_W = 2.03777\; {\rm GeV}, \qquad \Gamma_t=1.67432 {\rm GeV}.
\eea
We have neglected the QCD correction to the widths $\Gamma_W$ and
$\Gamma_t$, as no QCD corrections have been included in the one--loop
corrections to the $t\bar t$-pair production process.
The EW corrected widths of (\ref{params6}) are used in the calculation
of the cross sections that include the EW factorizable corrections. For the
calculation of the lowest order cross sections of (\ref{nmmn}) and
(\ref{nmud}) the corresponding lowest order SM values of the top quark and
$W$-boson widths are used.

Results for the total cross sections of reactions (\ref{nmmn}) and
(\ref{nmud}) at three different centre of mass (CMS) energies in the presence
of the following cuts on quantities $\delta_t, \delta_{\bar t},\delta_{W^+},
\delta_{W^-}$, defined in (\ref{deltas}),
\bea
\label{cuts}
\delta_t < 0.1, \quad \delta_{\bar t} < 0.1, \qquad
         \delta_{W^+} < 0.1, \quad \delta_{W^-} < 0.1,
\eea
are shown in Table 1. The second column shows the Born cross sections
calculated with the complete set of the lowest order Feynman diagrams.
The third column shows the Born `signal' cross section, {\em i.e.}
the cross section obtained with the two lowest order signal diagrams of
Fig.~\ref{fig:diags0}a only. We see that imposing the invariant mass cuts 
(\ref{cuts}) efficiently reduces the off resonance background, which 
becomes quite sizeable if the cuts are not imposed \cite{eett6f1,eett6f2}. 
The fourth and fifth
columns show the cross sections including the ISR and factorizable EW
corrections separately and the sixth column shows the results including both
the ISR and EW factorizable corrections. Note that the cross sections
of (\ref{nmud}) are almost exactly 3 times larger than the cross sections of
(\ref{nmmn}), in agreement with the naive counting of the colour degrees
of freedom. This is because the neutral current off resonance background
contributions that make
reaction (\ref{nmmn}) differ from (\ref{nmud}) are almost completely suppressed
in the presence of cuts (\ref{cuts}).

\begin{table}
{\small Table~1: Total cross sections of reactions (\ref{nmmn}) and
(\ref{nmud}) in fb at three different CMS energies in the presence of
cuts (\ref{cuts}).
The numbers in parenthesis show the uncertainty of the last decimals.}
\begin{center}
\begin{tabular}{|c|c|c|c|c|c|}
\hline
\hline
     \multicolumn{6}{|c|}{\rule{0mm}{7mm} $\eebnmbmn$} \\[1.5mm]
\hline
\hline
\rule{0mm}{7mm} $\sqrt{s}$ (GeV) & $\sigma_{\rm Born}$
       & $\sigma_{\rm Born}^{t^*\bar{t}^*}$
& $\sigma_{\rm Born+ISR}$ & $\sigma_{\rm Born+FEWC}$ & 
$\sigma_{\rm Born+ISR+FEWC}$ \\[2mm]
\hline
\rule{0mm}{7mm} 
  430 & 5.9117(54) & 5.8642(45) & 5.2919(91)& 5.6884(55) & 5.0978(53) \\[1.5mm]
  500 & 5.3094(50) & 5.2849(43) & 5.0997(51)& 4.9909(49) & 4.8085(48) \\[1.5mm]
 1000 & 1.6387(16) & 1.6369(15) & 1.8320(18)& 1.4243(14) & 1.6110(16) \\[1.5mm]
\hline
\hline
   \multicolumn{6}{|c|}{\rule{0mm}{7mm} $\eebnmbdu$} \\[1.5mm]
\hline
\hline
\rule{0mm}{7mm} $\sqrt{s}$ (GeV) & $\sigma_{\rm Born}$ 
       & $\sigma_{\rm Born}^{t^*\bar{t}^*}$ 
& $\sigma_{\rm Born+ISR}$ & $\sigma_{\rm Born+FEWC}$ &
$\sigma_{\rm Born+ISR+FEWC}$ \\[2mm]
\hline
\rule{0mm}{7mm} 
  430 & 17.727(16) & 17.592(13) & 15.857(20) & 17.052(16) 
                                             & 15.283(16) \\[1.5mm]
  500 & 15.950(15) & 15.855(13) & 15.311(15) & 14.977(16) 
                                             & 14.438(14) \\[1.5mm]
 1000 & 4.9134(48) & 4.9106(46) & 5.4949(55) & 4.2697(40) 
                                             & 4.8287(47) \\[1.5mm]
\hline
\end{tabular}
\end{center}
\end{table}

\begin{figure}[ht]
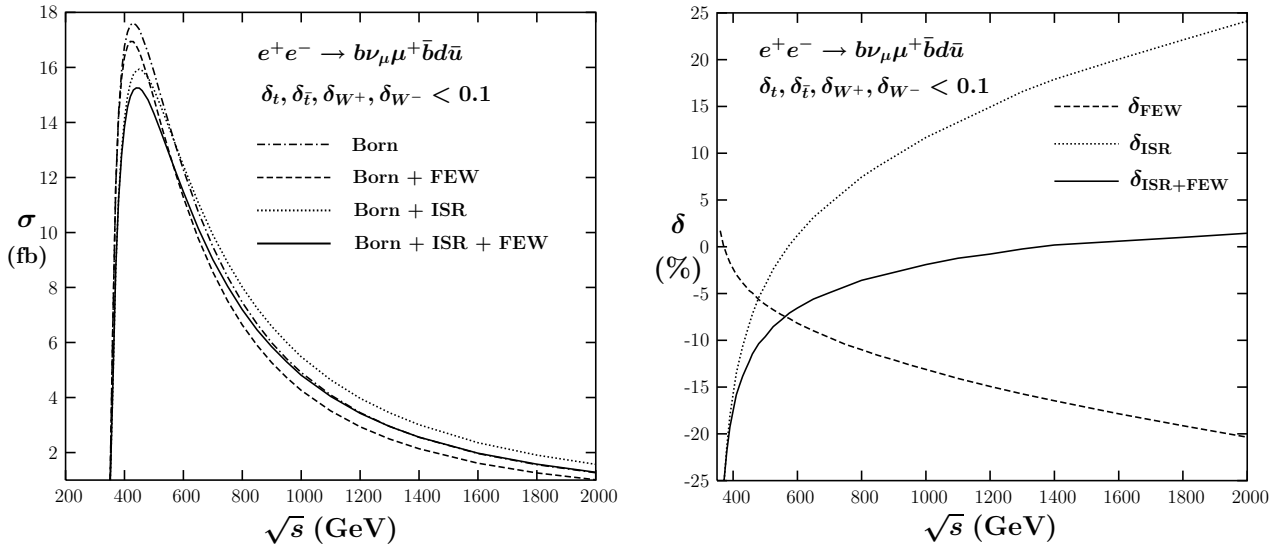

\begin{center}
\setlength{\unitlength}{1mm}
\begin{picture}(35,35)(60,-50)
\rput(5.3,-6){\scalebox{0.6 0.6}{\epsfbox{nmud.tt.epsi}}}
\end{picture}
\begin{picture}(35,35)(10,-50)
\rput(5.3,-6){\scalebox{0.6 0.6}{\epsfbox{rel.tt.epsi}}}
\end{picture}
\end{center}
\vspace*{4.0cm}
\caption{Total cross sections of (\ref{nmud}) including different classes of 
the SM radiative corrections (left) and corresponding relative corrections
(\ref{relcor})(right) as functions of the CMS energy.}
\label{fig3}
\end{figure}

How the radiative corrections for the six fermion reactions (\ref{ee6f})
depend on the CMS energy is illustrated in Fig.~\ref{fig3},
where, on the left hand side, we plot the total cross sections of reaction
(\ref{nmud}) as a function of the CMS energy, both in the lowest order
and including different classes of corrections.
The dashed-dotted line shows the Born cross section, the dotted line
is the cross section including the ISR correction, the dashed line shows
an effect of the factorizable EW correction while the solid line shows
an effect of the combined ISR and factorizable EW correction.
The plots on the right hand side of Fig.~\ref{fig3} show the corresponding
relative corrections 
\bea
\label{relcor}
\delta_{\rm cor.}&=&\frac{\sigma_{\rm Born+cor.}
-\sigma_{\rm Born}}{\sigma_{\rm Born}},
\qquad {\rm cor. = FEW, ISR, ISR+FEW.}
\eea
The dashed line shows the relative factorizable EW correction. The correction
is small and positive a few GeV above the $t\bar t$-pair production threshold,
but already about 20 GeV above the threshold it becomes negative
and it falls down logarithmically towards more and more negative values, due
to large logarithmic terms $\sim \left[\ln \left(m_W^2/s\right)\right]^2$
and $\sim \ln \left(m_W^2/s\right)$, reaching 20\% at $\sqrt{s}=2$~TeV. 
The dotted line shows the relative ISR correction, which on the other
hand is dominated by large collinear logarithms
$\left[\ln \left(s/m_e^2\right)\right]^2$
and $\ln \left(s/m_e^2\right)$. It starts from about $-25$\% at energies
close to the threshold and grows to almost $+25$\% at $\sqrt{s}=2$~TeV. 
Finally, the solid line shows the combined ISR and factorizable EW correction.
The net relative correction is dominated by the ISR: it is large and negative 
for energies not far above the threshold and it becomes
positive at high energies, reaching 1.4\% at at $\sqrt{s}=2$~TeV. 

\section{SUMMARY AND OUTLOOK}

We have calculated the SM predictions for top quark pair
production and decay into six fermions at a linear $e^+e^-$ collider.
We have included the factorizable EW $\cal{O}(\alpha)$ corrections in the
pole approximation and QED corrections due to the initial state radiation 
in the structure
function approach into SM predictions for the top quark pair 
production and decay into six fermions at the ILC. 
We have illustrated an effect of the radiative corrections on the predictions
by showing numerical results for two selected six-fermion reactions
(\ref{nmmn}) and (\ref{nmud}).
The ISR and factorizable EW radiative corrections are sizeable and therefore 
should be included
in the analysis of future precision data on the top quark pair production
and decay from the ILC.

In order to obtain a complete EW next to leading order result for six fermion
reactions (\ref{ee6f}) in the pole approximation one should include 
the nonfactorizable virtual photonic corrections
corresponding to an exchange of a virtual photon between the electrically charged
lines of the signal diagrams of Fig.~\ref{fig:diags0}(a) which has not been 
included in the shaded blobs of Fig.~\ref{fig:diags}. For example, an exchange of
a photon between the initial state electron and any of the final state fermions or
intermediate $W$ bosons, or between the $b$ and $\bar t$ quark, or its decay products
should be taken into account.
This would allow for inclusion of the real photon emission from the external legs in an 
exclusive way. Taking into account the QCD coreections
would also be higly desirable.

{\Large \bf Acknowledgements}

K.K. is grateful to the Alexander von Humboldt Foundation for supporting
his stay at DESY, Zeuthen, where this work has been partly done and to
the Theory Group of DESY, Zeuthen for kind hospitality.

\begin{appendix}

\section{PROJECTION OF MOMENTA}

In this appendix, we describe the projection procedure that has been 
used in order to associate each phase space point of the full 6-particle
phase space of reaction (\ref{ee6f}) with a point of the
reduced phase space of the on-shell top pair production
(\ref{eett}) and subsequent decay. The on-shell momenta $k_t$ and $k_{\bar t}$
of the $t$-quar and $\bar{t}$-antiquark, $k_{W^{\pm}}$ of the decay
$W^{\pm}$-bosons,
and $k_i$, $i=3,...,8$, of the decay fermions of reaction
(\ref{ee6f}) are constructed from the four momenta $p_i$, $i=3,...,8$,
of the final state fermions of reaction (\ref{ee6f}) with
the following projection procedure.

First the on-shell four momenta of $t$ and $\bar{t}$ in the
centre of mass system (CMS) are found in the following way

\parbox{8cm}{
\beas
\left|\vec{k}_{t}\right|&=&
\frac{\lambda^{\frac{1}{2}}\left(s,m_t^2,m_t^2\right)}{2s^{\frac{1}{2}}},\\
k^0_{t}&=&\left(\vec{k}_{t}^2 + m_t^2\right)^{\frac{1}{2}},
\eeas}
\hfill
\parbox{6cm}{
\beas
\vec{k}_{t}&=&\left|\vec{k}_t\right|
\frac{\vec{p}_3+\vec{p}_4+\vec{p}_5}
                 {\left|\vec{p}_3+\vec{p}_4+\vec{p}_5\right|}, \\
\vec{k}_{\bar{t}}&=&-\vec{k}_t, \quad
k^0_{\bar{t}}=\sqrt{s}-k^0_t.
\eeas}
\hfill
\parbox{2cm}{\bea \label{momtt}  \eea}

Then the four momenta $p_3$, $p_4$ and $p_5$ ($p_6$, $p_7$ and $p_8$)
are boosted to the rest frame of the $b f_1 \bar{f'_1}$
($\bar{b} f_2 \bar{f'_2}$) subsystem of reaction (\ref{ee6f}), where they
are denoted $p'_3$, $p'_4$ and $p'_5$
($p'_6$, $p'_7$ and $p'_8$). The projected four momentum
$k'_3$ of $b$ ($k'_6$ of $\bar b$) is determined in the rest frame
of $b f_1 \bar{f'_1}$ ($\bar{b} f_2 \bar{f'_2}$) according to
\bea
\left|\vec{k'}_{i}\right|=
\frac{\lambda^{\frac{1}{2}}\left(m_t^2,m_i^2,m_W^2\right)}{2m_t},
\qquad \vec{k'}_{i}=
\left|\vec{k'}_{i}\right|\frac{\vec{p'}_i}{\left|\vec{p'}_{i}\right|},
\qquad
{k'}_i^0=\left(\vec{k'}_{i}^2 + m_i^2\right)^{\frac{1}{2}}, \quad
i=3,6,
\label{k3k6}
\eea
which means that the directions of the $b$ and $\bar b$ momenta are kept unchanged
while their lengths are being altered.

The four momenta $p'_4$ and $p'_5$ ($p'_7$ and $p'_8$) are further boosted
to the rest frame of $f_1 \bar{f'_1}$ ($f_2 \bar{f'_2}$), where they
are denoted $p''_4$ and $p''_5$ ($p''_7$ and $p''_8$).
The projected four momenta $k''_4$ and $k''_5$ of $f_1$ and
$\bar{f'_1}$ ($k''_7$ and $k''_8$ of $f_2$ and $\bar{f'_2}$)
are in this frame determined according to
\bea
\left|\vec{k'}_{4}\right|\!\!\!&=&\!\!\!
\frac{\lambda^{\frac{1}{2}}\left(m_W^2,m_4^2,m_5^2\right)}{2m_W},
\qquad
\left|\vec{k'}_{7}\right|=
\frac{\lambda^{\frac{1}{2}}\left(m_W^2,m_7^2,m_8^2\right)}{2m_W}, \qquad
\vec{k''}_{i}=
\left|\vec{k''}_{i}\right|\frac{\vec{p''}_i}{\left|\vec{p''}_{i}\right|},
\quad i=4,7, \nn\\
\vec{k''}_{5}\!\!\!&=&\!\!\!-\vec{k''}_{4}, 
\qquad \vec{k''}_{8}=-\vec{k''}_{7}, \qquad\qquad
{k''}^0_{j}=\left(\vec{k''}_{j}^2 + m_j^2\right)^{\frac{1}{2}}, 
\quad j=4,5,7,8.
\label{k4578}
\eea
This again means that the directions of momenta of $f_1$, $\bar{f'_1}$,
$f_2$ and $\bar{f'_2}$ are kept unchanged while their lengths
are being altered.

The four momenta $k''_4$ and $k''_5$ ($k''_7$ and $k''_8$)
are now boosted to the rest frame of the on-shell $t$ ($\bar t$)
and, finally, $k'_3$, $k'_4$ and $k'_5$ ($k'_6$, $k'_7$ and $k'_8$)
are boosted from the $t$ ($\bar t$) rest frame to the CMS giving
the desired projected four momenta $k_i$, $i=3,...,8$.
As one can easily see from Eqs.~(\ref{momtt}--\ref{k4578}),
the projected momenta, except for satisfying the necessary on-shell relations
$k_i^2=m_i^2$, $i=3,...,8$,
fulfil also other required on-shell relations
\bea
\left(k_3+k_4+k_5\right)^2=\left(k_6+k_7+k_8\right)^2=
m_t^2, \qquad \left(k_4+k_5\right)^2=\left(k_7+k_8\right)^2=
m_W^2.
\eea

The described projection procedure is not unique. Moreover, it
strongly depends on the departures (\ref{deltas}) of invariant masses
$m_{345}$, $m_{678}$ of (\ref{mt}) from $m_t$, and of the invariant masses
$m_{45}$, $m_{78}$ of (\ref{mw}) from $m_W$.
How it works in practice is illustrated in Table~\ref{tabmom},
where two randomly selected sets of four momenta
$p_i$, $i=3,...,8$, $p_3+p_4+p_5$, $p_6+p_7+p_8$ , $p_4+p_5$,
$p_7+p_8$, and their projections $k_i$, $i=3,...,8$ , $k_t$,
$k_{\bar t}$ , $k_{W^+}$, $k_{W^-}$, respectively, are compared.
Momenta $p_i$ have been generated according to the Breit--Wigner
distribution in such a way that the invariant masses
of the $b f_1 \bar{f'_1}$, $\bar{b} f_2 \bar{f'_2}$,
$f_1 \bar{f'_1}$ and $f_2 \bar{f'_2}$ subsystems of reaction (\ref{ee6f})
fall into the vicinity of the masses of the corresponding primary on-shell
particles: $t$-quark, $\bar t$-antiquark, $W^+$- and $W^-$-boson, respectively.
\begin{table}
\label{tabmom}
{\small Table~A: A comparison of two randomly selected sets of the four momenta
         $p_i$, $i=3,...,8$, $p_3+p_4+p_5$, $p_6+p_7+p_8$ , $p_4+p_5$,
         $p_7+p_8$ and their projections $k_i$, $i=3,...,8$ , $k_t$, 
         $k_{\bar t}$ , $k_{W^+}$, $k_{W^-}$, respectively. Quantities
         $\delta_t=m_{345}/m_t-1$, $\delta_{\bar t}=m_{678}/m_t-1$, 
         $\delta_{W^+}=m_{45}/m_W-1$ and $\delta_{W^-}=m_{78}/m_W-1$ 
         describe relative departures of the corresponding final state
         particle subsystems from a mass-shell of the $t$, 
         $\bar t$, $W^+$ and $W^-$, respectively.}
{\small
\begin{center}
\begin{tabular}{|c|rrrr|rrrr|}
\hline
    & \multicolumn{4}{|c|}{\rule{0mm}{7mm}
$\delta_t=0.03\%$,  $\delta_{\bar t}=0.19\%$,} &
        \multicolumn{4}{c|}{$\delta_t=0.06\%$,  $\delta_{\bar t}=0.17\%$,}\\
GeV  & \multicolumn{4}{|c|}{$\delta_{W^+}=0.26\%$, $\delta_{W^-}=0.85\%$} &
      \multicolumn{4}{c|}{$\delta_{W^+}=0.78\%$, $\delta_{W^-}=3.23\%$}\\[2mm]
\hline
\hline
\rule{0mm}{7mm} 
$\!p_3$ &    154.0 &    141.4 &    --28.1 &     53.8 
      &    116.5 &     89.3 &     13.1 &     73.6\\
$k_3$ &    153.8 &    141.3 &    --28.0 &     53.8
      &    116.9 &     89.6 &     13.1 &     73.8\\[1mm]
$p_4$ &     22.9 &    --15.1 &     --7.7 &    --15.4
      &    117.6 &     92.8 &    --20.6 &    --69.3\\
$k_4$ &     22.9 &    --15.0 &     --7.8 &    --15.4
      &    117.4 &     92.5 &    --20.6 &    --69.2\\[1mm]
$p_5$ &     73.1 &     24.6 &     35.7 &     58.8
      &     16.0 &     --3.2 &      7.5 &     13.8\\
$k_5$ &     73.3 &     24.8 &     35.8 &     59.0
      &     15.7 &     --3.3 &      7.5 &     13.5\\[1mm]
$p_6$ &     64.5 &    --10.0 &     57.5 &    --27.0
      &    109.8 &    --78.5 &    --12.1 &    --75.7\\
$k_6$ &     64.0 &    --10.0 &     57.1 &    --26.9
      &    108.1 &    --77.2 &    --11.9 &    --74.6\\[1mm]
$p_7$ &    112.1 &   --109.2 &    --20.7 &    --15.1
      &    106.1 &    --95.4 &     33.7 &     32.0\\
$k_7$ &    112.4 &   --109.5 &    --20.4 &    --15.1
      &    107.8 &    --97.2 &     34.5 &     31.4\\[1mm]
$p_8$ &     73.5 &    --31.7 &    --36.8 &    --55.1
      &     33.9 &     --4.9 &    --21.6 &     25.7\\
$k_8$ &     73.6 &    --31.5 &    --36.7 &    --55.5
      &     34.1 &     --4.4 &    --22.6 &     25.1\\[1mm]
$p_3+p_4+p_5$ &    249.9 &    150.9 &      0.0 &     97.3
              &    250.1 &    178.9 &      0.0 &     18.0\\
$k_t$ &      250.0 &    151.0 &      0.0 &     97.4
      &      250.0 &    178.8 &      0.0 &     18.0\\[1mm]
$p_6+p_7+p_8$ &    250.1 &   --150.9 &      0.0 &    --97.3
              &    249.9 &   --178.9 &      0.0 &    --18.0\\
$k_{\bar t}$ &     250.0 &   --151.0 &      0.0 &    --97.4
             &     250.0 &   --178.8 &      0.0 &    --18.0\\[1mm]
$p_4+p_5$ &     95.9 &      9.5 &     28.1 &     43.5
          &    133.6 &     89.6 &    --13.1 &    --55.5\\
$k_{W^+}$ &     96.2 &      9.8 &     28.0 &     43.6
          &    133.1 &     89.2 &    --13.1 &    --55.8\\[1mm]
$p_7+p_8$ &    185.6 &   --140.9 &    --57.5 &    --70.3
          &    140.0 &   --100.4 &     12.1 &     57.7\\
$k_{W^-}$ &    186.0 &   --141.0 &    --57.1 &    --70.5
          &    141.9 &   --101.6 &     11.9 &     56.6\\[1mm]
\hline
\end{tabular}
\end{center}}
\end{table}

\end{appendix}

\end{document}